\documentclass[namedreferences]{kluwer}
\begin{document}
\begin{opening} 

\title{The Evolution of Stellar Populations}
\subtitle{Discussion Session}
\author{Angeles I. \surname{D\'\i az}}
\institute{Departamento de F\'{\i}sica Te\'{o}rica, C-XI,
  Universidad Aut\'{o}noma de Madrid, 28049 Madrid, Spain}
\author{Eduardo \surname{Hardy}\footnote[1]{The National Radio Astronomy Observatory is operated by Associated Universities, Inc., under a cooperative agreement with the National Science Foundation.}}
\institute{National Radio Astronomy Observatory (NRAO) \\ Casilla 36-D, Santiago, Chile }

\runningauthor{A. D\'\i az \& E. Hardy}
\runningtitle{Stellar Populations}



\begin{abstract}

We summarize the discussion section on ``Evolution of Stellar Populations'' we led on May 27, 2000 in Granada, Spain, as part of the Euroconference on {\it The Evolution of Galaxies. I- Observational Clues}. The discussion was organized around two groups of topics. In the first, {\it Population Synthesis}, the accent was partially placed on the use of tools and techniques centered around the question of the unicity of the models, their sensitivity to input and the question of the age-metallicity degeneracy. In the second group, {\it Stellar Systems} a stronger accent was placed on astrophysical questions, although we included there the need for ``truth tests'' that apply spectral synthesis techniques to objects for which there is detailed {\it a priori} knowledge of their stellar populations. We also provide a partial comparison between the present knowledge of these topics and that which existed at the time of the Crete Conference of 1995.
\end{abstract}

\end{opening} 

\section{Introduction}

Five years ago, in October 1995, a conference was held in Crete which brought together in a friendly environment observers and theoreticians working in the field of galaxy evolution. A substantial part of that meeting was devoted to the study of the stellar populations of galaxies and its different analytical and observational tools. The overall results were summarized, as was also the case in this conference, by John Huchra \cite{Leitherer96}. The Crete conference established a reference baseline for what was known about our subject at that time, and a natural starting point to evaluate progress since then. The reader is invited to examine the excellent summary mentioned above and compare it with the present one with a question in mind: What has changed since Crete? 

At Crete it was recognized that the existing stellar libraries were good only over a limited spectral and chemical abundance range and, moreover, that much better physical calibrations were needed. It was also recognized that complete sets of stellar models for a wide range of metallicities and masses, including mass loss and rotation, should be computed. The importance of performing ``truth tests", that is of confronting the results of population synthesis models based on the spectral properties of the integrated light of galaxies with the observations of globular clusters (GC) and Local Group (LG) dwarf galaxies was stressed, as it was the need for compiling good quality multi-wavelength observations of selected objects. The latter would become the basis of a reference library. The former would show the limitation of spectral synthesis, by nature an ``inverse problem'', when confronted with the age-metallicity results issued from the presumably more reliable deep color-magnitude diagrams of the same object. Quantifying internal errors both in the theoretical models and in the observational data was also deemed essential.
 
At the end of the Crete meeting expectations ran high for new studies based on the unprecedented resolution and depth provided by both Hubble Space Telescope (HST) and the new generation of ground based telescopes: color-magnitude diagrams (CMD) for galaxy populations, Wolf-Rayet content, upper mass of the Initial Mass Function (IMF), etc. The then new HST images of Eta Carina, NGC~2403, and IZw18 became the ``stars of the show", and many of us were impressed by the identification of young (1 to 5 Myr) bursts of star formation in the latter object, one of the all-time favorite astronomical playgrounds.

Following this lead we suggested the audience at the Granada conference an outline for the discussion session based on the following very general and loose scheme as ordered by category and by subject:

\begin{itemize}
\item{\bf A. Population Synthesis (partial accent on tools)}
\begin{itemize}
\item{Unicity of models}
\item{Sensitivity to input (IMF, stellar libraries, evolutionary tracks)}
\item{Is the age-metallicity degeneracy still with us? If so, can it be beaten?}
\end{itemize}
\item{\bf B. Stellar Systems (partial accent on astrophysical questions)}
\begin{itemize}
\item{``Truth Tests''. Globular clusters, nearby galaxies, the galactic bulge: how well can we date GCs via integrated light; how well can we solve the age-metallicity degeneracy via integrated light?}
\item{Giant Ellipticals: are there ``young'' stars there; if so, how metal--rich are they?}
\item{Dwarf Irr, Blue Compact Galaxies (BCD's), starbursts {\it `\`a la} IZw18: are there ``old'' stars there; if so, how metal-poor are they?}
\end{itemize}
\end{itemize}

The ensuing discussion was quite lively and a large number of people took part in it. We are aware that we have probably not done justice to all that was said, nor to all who participated. Then not all of the subjects proposed were discussed, nor all of them were discussed in the same depth, nor recorded by us with the same degree of detail. References to the many posters related to our session were interspaced within the text where appropiate, and the reader will find a full description of each of them in this volume.

\section{Population Synthesis}

\subsection{Models: libraries, evolutionary tracks, errors, IMF}

G. Bruzual, who had given the conference's review paper on models, insisted upon the fact that spectral libraries of intermediate resolution ($\sim 10\ \AA$) providing a better coverage of the metallicity and metallicity ratios were required and not yet satisfactory. He reminded us that in the UV we are still using old data for about 40 stars taken by IUE. More, better and more modern data are required, and not only in the UV but also in the IR. These libraries can be supplemented by or be a complement to existing cluster libraries such as those of Bica and Alloin (\opencite{Bica87}). Along the same lines, A. Vazdekis commented that the spectral resolution of the Lick system, for which an extensive library exists, is adequate for the velocity dispersion of most galaxies, but pointed out that there is a lack in that library of hot stars and of low-metallicity stars. In addition he feels a better flux calibration is needed.

The difficulties in using evolutionary models for Red Giant Branch (RGB) and 
Asympthotic Giant Branch (AGB) stars were stressed by D. Schaerer who questioned the existence of an adequate observational sample for these stars. Indeed, is there a proper modeling of AGB stars in Single Stellar Populations (SSP) (A. Bressan)? How reliably can we use Wolf Rayet features in galaxies to date ionizing stellar populations (A. Maeder, D. Schaerer, D. Kunth)? Again, is the effect of binaries properly taken into account (M. Cervi\~no)?

The issue of unicity of the models was raised, but did not unduly stirred the audience. D. Alloin, a practitioner of the art, did not regard this as a major drawback when some astrophysical assumptions were introduced, such as reasonable tracks in the age-metallicity plane: old populations should be more metal-poor than young ones, etc. M. Tosi stressed the need for studies of good Color Magnitude Diagramas (CMDs) of galaxies to test the models.

Are we happy with stellar evolutionary tracks computed only, or mainly, with solar abundance {\it ratios}? It would seem that scaling schemes to apply solar-ratio tracks to complex situations are well in hand, and this was also not regarded as a major drawback. 

G. Stasi\'nska raised the issue of the impact of binaries in synthesis models. R. Schulte-Ladbeck estimates at about 30 \% the fraction of binaries needed to synthesize open clusters. As far as quantifying theoretical uncertainties is concerned, nothing was proposed beyond recommending the use of different evolutionary tracks for comparison purposes.

The issue of the universality of the IMF was raised, but no consensus was reached. J. Melnick, who had presented a theoretical approach to the IMF formation in an earlier talk, specified during the present discussion that his work is restricted to starbursts, and will not necessarily apply, for example, to the Magellanic Clouds (MCs) where P. Massey finds IMF shapes which are at variance with the classical function.

Some but not much evolution is detectable in this area since Crete.

\subsection{Models:Age {\em vs} Metallicity Degeneracy: an old problem. New solutions? }

How to break the age-metallicity degeneracy that has been with us for so long? Two general approaches were proposed, one based on the study of the CMD, the other based on the study of spectral indices.

C. Gallart builds CMDs reaching the old Main Sequence (MS) turnoffs and with good statistics in all phases of evolution. She insists that in this way detailed information on the age distribution of the stars is gained from their distribution along the MS and subgiant branch. Once this distribution is determined, the possible metallicity distributions obtained from a model fit to the detailed distribution of stars in the CMD are strongly constrained and are relatively unique. 

A lively discussion about methods and results took place among some of the main practitioners of CMD synthesis including C. Gallart, E. Grebel and M. Tosi. E. Grebel suggested using deep (HST) CMDs of open clusters as input for synthesis models. In this way open clusters of known age and metallicity are used as building blocks for the CMD diagram of complex systems. Something similar is in fact done in spectral synthesis (e.g. \opencite{Bica87}).

A number of participants, including A. Bressan, C. Maraston and  A. Vazdekis, discussed the use of spectral indices to break the age-metallicity degeneracy. One of the most intriguing recent results in this area is the unacceptably old age of 20 Gyr for 47Tuc obtained using the high-resolution Balmer index H$\gamma$$_{HR}$\ (\opencite{Gibson99}). How are we to interpret this result? C. Maraston proposed an answer (\opencite{Maraston2000}) which has important consequences for the dating of elliptical galaxies: H$\beta$ (and presumably other Balmer indices) increases steeply for metal-poor globular clusters in such a way that a population containing even a small fraction of old metal-poor stars will produce strong Balmer lines. Using their calibration of H$\beta$ based on new SSP models they find an age of 15 Gyr for 47Tuc consistent with most CMD studies. Furthermore, they argue that there's no need for young stars in elliptical galaxies.

In the end, the question is whether a ``smoking gun'' index, essentially a Balmer line (thus sensitive to main sequence turn-off position), versus a (mostly) metallicity index such as those proposed by \inlinecite{Worthey94} or the CaT as calibrated by \inlinecite{Molla2000} for example (see also the poster contribution by J. Gorgas et al.), will break the degeneracy. D. Alloin proposes the alternative use of a large number of indices which are individually sensitive to both parameters but which together can provide the required degeneracy breaking.

Is the problem solved, or not solved at all? Stay tuned.

\section{Stellar Systems}

\subsection{``Truth tests''}

One of the outstanding issues, which has only seen partial progress since Crete, is that of testing spectral synthesis on objects for which the age-metallicity distribution is known {\it a priori}. Astronomers tend to believe color-magnitude diagrams better than they believe spectral synthesis: will the twain ever meet? The obvious objects in which to start this enterprise are globular clusters as they are single-age, single-metallicity objects. We have seen above that conflicting results have been obtained for the classical test-bench, 47Tuc. But at least in this latter case it seems that proper calibration of the spectral indices may give consistent results. The situation for the more interesting (and needed!) composite populations of galaxies is still uncertain. E. Hardy advocated (as he did in Crete) the study of nearby resolved populations such as the galactic bulge, and the Magellanic Clouds. Observations of the MCs are underway in collaboration with C. Gallart, D. Alloin and others. In this respect the last few years have witnessed a certain loss of innocence in dealing with the CMDs of composite systems. The question is what {\it a priori} means in the first sentence of this section. CMDs of composite system do not unveil their secrets just by visual inspection. CMD synthesis work by C. Gallart, E. Tolstoy, E. Grebel and others has shown how difficult it is to disentangle ages from metallicity in the CMD of these systems (see above), and has shown that a proper analysis of a CMD requires careful modeling. This points towards a limitation of the proposed ``truth tests'' in that the CMD itself does not yield an answer which is model-independent. Both synthesis methods are indeed model-dependent. It is the consistency between their results, and the way in which they might complement each other that is now the real issue to be resolved before spectral synthesis can confidently jump beyond the Local Group.

\subsection{On the existence of old populations in {\em all} dwarf irregulars and Blue Compact
Galaxies}

For some in the audience, if there has been something new and exciting since Crete it is probably the convincing observational evidence for the presence of an old stellar population in the strongest candidate to be a primeval galaxy: IZw18. In fact, given the leading role of this particular galaxy during the whole session, it was generally felt that a conference devoted solely to IZw18, the most metal poor blue compact (HII) galaxy known, would be welcome. The evidence comes from  separate analyses of HST data (WFPC2 (\opencite{ATG99}) and NICMOS (\opencite{Ost00})) showing the presence of AGB stars corresponding to a stellar population of at least 1 Gyr, a result which is not seriously compromised even if the distance to the object is significantly larger than assumed. Moreover, IZw18 does not seem to be unique amongst its class. There seems to be general agreement that there is evidence for the presence of an old (several Gyr) population in BCGs, as shown by  U. Hopp from HST/NICMOS data of a sample of these objects (see also the poster contribution by R. Schulte-Ladbeck et al.). However, a similar consensus has not yet been reached regarding the way in which star formation proceeds in these galaxies. Is it going on in short bursts separated by long periods of laziness as generally accepted? Or there is a continuous, albeit low, star formation rate on top of which star bursts occur as proposed by F. Legrand?

Another related question is whether BCG can in some way be identified with the high redshift dwarfs which would be experiencing their first star formation episodes. Can this hypothesis be tested? The analysis of the rest UV spectrum of MS1512-cB58 (z=2.73), conveniently amplified by a gravitational lens, by Sally Heap et al. points to a constant SFR for this galaxy, otherwise a starburst-like galaxy. On the other hand, the local star forming galaxies (UCM sample) do not seem to be composed of very spectacular objects, but rather by objects with masses around 5$\times$10$^{\rm 10}$ M$_{\odot}$ undergoing a starburst involving about 2\% of their mass (see the poster contribution
by A. Gil de Paz et al.).

\subsection{New windows for the study of stellar populations}

Due to fast progress in observing and analyzing high redshifts objects we will be obtaining a large amount of data in the IR (near to far). This means observing in the IR the UV rest frame spectra of the high {\em z} objects. On the other hand, recent advances in instrumentation make possible the observation of spectral properties of stellar populations in the IR. These IR studies are already yielding very interesting results, as can be appreciated by looking at the numerous poster contributions on this topic: the underlying populations of BCG (mentioned above); stellar populations in AGN (through the H window); composite stellar systems (CaT) kinematics; edge-on galaxies; inner Milky Way Galaxy (MWG) and so on and so forth. As a result, we need to study in the UV rest frame stellar populations of galaxies which will be observed in the near IR for objects at {\em z} = 3. Not much seems to have been done in this direction (but see \opencite{aGD97} and \opencite{bGD97}).

In a related way, but moving this time towards the mm and sub-mm spectral range, the fast approaching ALMA era will create unheard of opportunities and challenges for the study of the early Universe. Preparatory work on nearby objects at ``shorter'' wavelengths will also be required, and it was gratifying that some in the audience showed an appreciation for this opportunity.

The multiwavelength approach is undoubtedly the most complete way of studying stellar populations as well as to test population synthesis models. An example of this approach was presented by R. Gonz\'alez Delgado for the giant HII region NGC~604. Multivawelength observations, on the other hand, introduce extra observational difficulties and, as A. D\'\i az pointed out, it might be worth to have some representative objects with complete wavelength coverage, at least from the UV to the far IR. In some cases this would involve completing and extending existing data libraries. Hopefully we could all work together to generate a database that could serve to test our stellar population synthesis models.

\section{Conclusion and desiderata: on the way to the next Euroconference}

Let us take again, for the purpose of summarizing this discussion and preparing the list of problems we would like to see solved in the near future, the discussion scheme proposed in the Introduction:

\begin{itemize}
\item{\bf A. Population Synthesis}
\begin{itemize}
\item{{\bf Unicity of models}: not much emphasis was placed on this and for some it simply is a non-issue. It may however be a dormant problem which may raise its head as we come closer to testing the consistency of models based on spectral synthesis and color magnitude diagrams.}
\item{{\bf Sensitivity to input} (IMF, stellar libraries, evolutionary tracks). The universality of the IMF remains an open fundamental question especially for evolutionary synthesis where it is one of the fundamental input parameters. Also, we certainly need to improve on the present spectral libraries, not only for stars, but also for key reference objects. The improvement should come from better physical calibrations, a more complete coverage of the metallicity range, and the extension of the wavelength coverage. Improvement in the quality and scope of stellar evolutionary tracks, and the required atmospheres, is likely to come in the near future. We hope to see, at least, a better agreement between theory and observations in the all-important red giant branch as well as a more detailed treatment of non-solar abundance ratios}

\item{{\bf Is the age-metallicity degeneracy still with us? If so, can it be beaten?}The answer to both question is a qualified {\it yes}. On the one hand some of the spectral tools used on simple systems, such as globular clusters, give contradictory results, but on the other hand more realistic SSP models are possible and they seem to lead towards consistency with the ages determined via the color magnitude diagram. For composite systems containing a range of ages and metallicities, the same spectral tools, once properly calibrated, may lift the degeneracy. Synthetic CMD for complex systems are coming of age and they are likely to present their own set of new problems.}
\end{itemize}
\newpage
\item{\bf B. Stellar Systems (partial accent on astrophysical questions)}
\begin{itemize}
\item{`{\bf `Truth Tests''}. Globular clusters, nearby galaxies, the galactic bulge: how well can we date GCs via integrated light; how well can we solve the age-metallicity degeneracy via integrated light? Little has been done on this subject, but observations of key LG members are on the way. Equally importantly, we now have a better idea of how to interpret CMD of complex non-coeval systems.}

\item{{\bf Giant Ellipticals}: are there ``young'' stars there; if so, how metal--rich are they? The jury is still out. Conventional wisdom used to dictate that ellipticals were old, metal-rich, basically single-burst objects. Then conventional wisdom evolved and intermediate-age stars appeared more likely. Now, coming full circle, we are told that we can reproduce the strong Balmer lines in ellipticals with even a small contribution from old metal-poor stars. Are we reaching a climax here?}

\item{{\bf Dwarf Irr, BCD's, starbursts {\it \`a la} IZw18}: are there ``old'' stars there; if so,
how metal-poor are they? Old stars seem to be everywhere, unless we find a true primordial galaxy. But IZw18 does not seem to be the one. In any case the history of star formation in these galaxies is largely unknown. In all of this ALMA is likely to tell us much, and we can hardly wait.}
\end{itemize}
\end{itemize}

Let's hope that by the end of this series of Euroconferences many of the questions raised here will have found an answer. Many more new questions are waiting for us.

\begin{acknowledgements}

We would like to thank Elena Terlevich who faithfully took note of what was said during the discussion session thus providing the basis for this written contribution.

\end{acknowledgements}

{}

\begin{thebibliography}{}

\bibitem[\protect\citeauthoryear{Alloisi et al.}{1999}]{ATG99}
Aloisi, A., Tosi, M. and Greggio, L., 1999, 
{\it AJ} {\bf 118}, 
~302

\bibitem[\protect\citeauthoryear{Bica \& Alloin}{1987}]{Bica87}
Bica, E., and Alloin, D., 1987, 
{\it A\&A\/} {\bf 186}, 
~49

\bibitem[\protect\citeauthoryear{Gibson et al.}{1999}]{Gibson99}
Gibson, B. K., Madgwick, D. S., Jones, L. A., Da Costa, G. S., and Norris, J. 
E., 1999, 
{\it AJ\/} {\bf ~118}, 
~1268

\bibitem[\protect\citeauthoryear{Gonz\'alez Delgado et al.}{1997a}]{aGD97}
Gonz\'alez Delgado, R.M., Leitherer, C. and Heckman, T., 1997,
{\it ApJ} {\bf 489}, 
~601

\bibitem[\protect\citeauthoryear{Gonz\'alez Delgado et al.}{1997b}]{bGD97}
Gonz\'alez Delgado, R.M., Leitherer, C., Heckman, T. and Cervi\~no, M., 1997, {\it ApJ} {\bf 483},
~705

\bibitem[\protect\citeauthoryear{ASP Conferences Series, vol.98}{1996}]{Leitherer96}Leitherer, C.,
Fritze-von Alvensleben, U., and Huchra, J.. 1996,
From Stars To Galaxies: The Impact Of Stellar
Physics On Galaxy Evolution. 
{\it ASP Conference Series\/}, 
{\bf Vol. ~98}, 
pp.~597--~609.

\bibitem[\protect\citeauthoryear{Maraston \& Thomas}{2000}]{Maraston2000}
Maraston, C., and Thomas, D.,2000, 
{\it astro-ph/0004145} 

\bibitem[\protect\citeauthoryear{Moll\'a \& Garc\'{\i}a--Vargas} {2000}]{Molla2000}
Moll\'a, M., and Garc\'{\i}a--Vargas, M. L., 2000, 
{\it A\&A\/} {\bf 358}, 
~18

\bibitem[\protect\citeauthoryear{Ostlin}{2000}]{Ost00} Ostlin, G., 2000,
{\it ApJ} {\bf 535}, 
~L99.

\bibitem[\protect\citeauthoryear{Worthey}{1994}]{Worthey94}
Worthey, G., 1994,
{\it ApJS\/} {\bf95}, 
~107

\end{thebibliography}
\end{document}